\documentclass[preprint]{elsarticle}
\usepackage{amsmath}
\usepackage{amssymb}
\usepackage{amsthm}
\usepackage{psfrag}
\usepackage{graphicx}
\usepackage{color}
\usepackage{upgreek}
\journal{Physics Letters B}
\begin{document}
\begin{frontmatter}
\title{ Darkness without dark matter and energy -- generalized unimodular gravity}

\author{A. O. Barvinsky}
\address{Theory Department, Lebedev Physics Institute, Leninsky Prospect 53, Moscow 119991, Russia\\
Tomsk State University, Department of Physics, Lenin Ave. 36, Tomsk 634050, Russia}\ead{barvin@td.lpi.ru}
\author{A. Yu. Kamenshchik}
\address{Dipartimento di Fisica e Astronomia, Universit\`a di Bologna and INFN, Via Irnerio 46,40126 Bologna,
Italy\\
L.D. Landau Institute for Theoretical Physics of the Russian
Academy of Sciences, Kosygin str. 2, 119334 Moscow, Russia}\ead
{kamenshchik@bo.infn.it}

\begin{abstract}
We suggest a Lorentz non-invariant generalization of the unimodular gravity theory, which is classically equivalent to general relativity with a locally inert (devoid of local degrees of freedom) perfect fluid having an equation of state with a constant parameter $w$. For the range of $w$ near $-1$ this dark fluid can play the role of dark energy, while for $w=0$ this dark dust admits spatial inhomogeneities and can be interpreted as dark matter. We discuss possible implications of this model in the cosmological initial conditions problem. In particular, this is the extension of known microcanonical density matrix predictions for the initial quantum state of the closed cosmology to the case of spatially open Universe, based on the imitation of the spatial curvature by the dark fluid density. We also briefly discuss quantization of this model necessarily involving the method of gauge systems with reducible constraints and the effect of this method on the treatment of recently
  suggested mechanism of vacuum energy sequestering.
\end{abstract}
\end{frontmatter}

\section{Introduction}

Dark matter and dark energy phenomena form a dark side of modern precision cosmology and, therefore, represent an unprecedentedly rich playground for various modifications of general relativity (GR). Perhaps, conceptually the most interesting versions of these modifications are the ones which do not involve special types of gravitating matter and originate from the purely metric sector of the theory, like local $f(R)$-gravity or nonlocal cosmology models \cite{f(R), Deser-Woodard}. Usually such modifications are equivalent to adding or removing some local degrees of freedom. Even more interesting is the case when a nontrivial modification occurs without changing the balance of local physical variables -- darkness arises without dark energy or dark matter constituents. Known examples of such a concept include, in particular, the unimodular (UM) gravity \cite{unimodular,unimodularHT,unimodular1}, the theory of vacuum energy sequestering \cite{sequester,sequester1}, QCD holonomy mechanism of dark energy \cite{QCDholonomy} and others. Unimodular gravity differs from the Einstein GR by the requirement that at the kinematical level the full set of metric coefficients is subject to the restriction of the unit determinant of the metric tensor. Rather anti-intuitive conclusion that this theory has the same number of local degrees of freedom as GR \cite{UMcan} can be explained by the fact that reduction in the number of independent field variables is compensated by the reduction of the local gauge invariance group, and the main effect of the unimodular modification is the origin of one global degree of freedom playing the role of the cosmological constant.

Extension of the physical sector of the theory by a partial violation of gauge invariance is a well-known and rather popular phenomenon. In particular, reduction from Lorentz symmetry to anisotropic scaling invariance in Lifshitz models is very productive in condensed matter theory context \cite{Lifshitz}, while a similar modification in Horava gravity models \cite{HGrav} opens prospects for renormalizable unitarity preserving gravity theories. Other examples can be found in \cite{Venturi,Akhmeteli}. Here we will consider the synthesis of Lorentz violation with the concept of unimodular gravity \cite{unimodular,unimodularHT,unimodular1}. This generalized UM gravity incorporates Lorentz violation in the definition of the reduced configuration space of metric coefficients -- instead of the requirement of a unit metric determinant this theory is based on the metric field satisfying the following constraint
    \begin{eqnarray}
    N=N(\gamma),\quad \gamma\equiv{\rm det}\,\gamma_{ij},             \label{class}
    \end{eqnarray}
where $N=(-g^{00})^{-1/2}$ is the lapse function and $N(\gamma)$ is some function of $\gamma$ -- the determinant of the spatial metric $\gamma_{ij}$ in the ADM $(3+1)$-decomposition of metric coefficients $g_{\mu\nu}$,
    \begin{eqnarray}
    g_{\mu\nu}dx^\mu dx^\nu =
    (N_i\,N^i-N^2)\,dt^2+2N_i\,dt\,dx^i
    +\gamma_{ij}dx^i\,dx^j.               \label{4metric}
    \end{eqnarray}
Here $x^\mu=(t,x^i)$, $\mu=0,1,2,3$, $i=1,2,3$ and $N_i=g_{0i}$ is the correponding shift function.

The motivation for such a generalization of the unimodular gravity is as follows. To begin with, the class of metrics subject to (\ref{class}) includes the original unimodular theory corresponding to $N(\gamma)=1/\sqrt\gamma$. The right hand side of (\ref{class}) is invariant under spatial rotations, so that this is a minimal breakdown of Lorentz symmetry from $O(1,3)$ to $O(3)$. Another reason to consider it is an interesting fact that at the classical level such a theory effectively incorporates a special type of matter source -- dark fluid with a nonlinear (general barotropic) equation of state. Thus it goes beyond a conventional unimodular gravity by generating the perfect fluid characterized not by just vacuum energy with $p=-\varepsilon$, but by a nontrivial pressure as well. Finally, for a simple class of power-like functions $N(\gamma)$ in (\ref{class}) it generates an equation of state $p=w\varepsilon$ with a constant $w$ and, moreover, in the comoving reference frame of this fluid renders the density and pressure constant both in space and time.\footnote{Since this model violates general coordinate invariance this property of density and pressure becomes frame dependent.} Thus, similarly to the original unimodular gravity it can incorporate as a spacetime constant of motion the analogue of dark energy which is free from clustering but has a constant polytropic parameter $w$ different from $-1$. In the particular case of a pressureless dust with $w=0$, corresponding to $N(\gamma)={\rm const}$, the density of this dust is characterized by a single function of spatial coordinates entirely fixed by the initial conditions, which can be interpreted as a model of inhomogeneous distribution of dark matter similar to the mechanism of mimetic model \cite{mimetic}.

Here we analyze this model at the classical level and show that on shell (without extra matter sources) it is equivalent to general relativity with this special type of perfect fluid. Its ``darkness" can be intuitively interpreted  as the absence of {\em local} degrees of freedom  of this fluid, and its effective manifestation can in principle be either the dark energy or dark matter. Rigorous counting its degrees of freedom, which is important for the quantization of this model, requires the analysis of its local gauge invariance. Usual diffeomorphism invariance is obviously broken by the restriction (\ref{class}) on metric coefficients, which leads to a preferred spacetime foliation by spacelike hypersurfaces. However, there exist reduced diffeomorphisms which leave the theory locally gauge invariant and turn out to be a generalization of volume preserving diffeomorphisms of the unimodular gravity. We briefly discuss them and show that their origin naturally leads to the theory with reducible (linearly dependent) generators. At the quantum level it is subject to Batalin-Vilkovisky technique \cite{BV} which allows one to quantize the theory without explicitly disentangling its physical sector.

We conclude the paper by the discussion of how this model can be used within the initial conditions problem in cosmology. Dark fluid of generalized UM gravity can be used to imitate the effect of spatial curvature. This might extend the predictions of the cosmological density matrix construction \cite{SLIH}, which are valid only in the spatially closed model, to the phenomenologically more preferable open model with flat space foliation. Another potential application could be the mechanism of sequestering the back reaction effect of quantum vacuum energy recently suggested as a possible solution of Planckian hierarchy and cosmological constant problems \cite{sequester,sequester1}. Remarkably, the method of careful treatment of the global physical mode responsible for the locally inert dark fluid is the same as that of the sequestering mechanism -- the canonical version of the BV method \cite{BV}, which might clarify acausality puzzles of this mechanism and extend it to noncompact spacetimes.

\section{Dark fluid and its generalized unimodular invariance}
The simplest way to handle the constraint (\ref{class}) on metric coefficients is not to explicitly substitute it in the Einstein action, but rather incorporate it into the action with the Lagrange multiplier $\lambda$,
    \begin{eqnarray}
    S=\int d^4x\,\left\{\frac{M_P^2}2\,g^{1/2}R(g)
    -\lambda\left(\frac1{\sqrt{-g^{00}}}
    -N(\gamma)\right)\right\}.               \label{action}
    \end{eqnarray}
Varying this action with respect to $\lambda$ and $g_{\mu\nu}$ one obtains the restriction (\ref{class}) on the metric and the Einstein equation with the perfect fluid matter stress tensor
    \begin{eqnarray}
    &&R^{\mu\nu}-\frac12\,g^{\mu\nu}R=\frac1{M_P^2}\,T^{\mu\nu},\\
    &&T^{\mu\nu}\equiv-\frac2{g^{1/2}}\frac\delta{\delta g_{\mu\nu}}\!\int d^4x\,\lambda\left(\!\frac1{\sqrt{-g^{00}}}
    -N(\gamma)\!\right)=\varepsilon\,u^\mu u^\nu
    +p\big(g^{\mu\nu}+u^\mu u^\nu\big),       \label{stresstensor}
    \end{eqnarray}
where the four-velocity $u^\mu=-g^{\mu 0}N$ is a future pointing vector normal to spacelike hypersurfaces of the ADM foliation (\ref{4metric}), and its energy density and pressure read
    \begin{eqnarray}
    \varepsilon=\frac\lambda{2\sqrt\gamma},\quad
    p=\frac\lambda{\sqrt\gamma}
    \left(\frac\gamma{N}\frac{dN}{d\gamma}\right). \label{denpress}
    \end{eqnarray}
Thus, this dark fluid satisfies the equation of state $p=w\varepsilon$ with a generally nonconstant parameter $w=w(\gamma)$ given by
    \begin{eqnarray}
    w=2\,\frac\gamma{N}\frac{dN}{d\gamma}
    =2\,\frac{d\ln N}{d\ln\gamma}.
    \end{eqnarray}

Similarly to the UM gravity \cite{unimodular} the generalized unimodularity condition (\ref{class}) is not invariant under generic diffeomorphisms of the metric -- Lie derivatives with respect to the 4-dimensional vector field $\xi^\mu$ which in the $(3+1)$-decomposition can be written down as a column,
   \begin{eqnarray}
    \delta_\xi g^{\mu\nu}=-\nabla^\mu\xi^\nu-\nabla^\nu\xi^\mu,\quad \xi^\mu=
    \left[\begin{array}{c}
    \,\xi^0\,\\
    \,\xi^i\,
    \end{array}\right].
    \end{eqnarray}
However, this condition remains invariant under reduced diffeomorphisms with respect to the subset of vector fields $\mbox{\boldmath$\xi$}^\mu$ satisfying the equation
    \begin{eqnarray}
    \delta_{\mbox{\boldmath$\xi$}}\big(N-N(\gamma)\big)\,\Big|_{\;N=N(\gamma)}=
    N\big[\,\partial_t\mbox{\boldmath$\xi$}^0
    -(1+w)\,N^i\partial_i\mbox{\boldmath$\xi$}^0
    -w\,\partial_i\mbox{\boldmath$\xi$}^i\,\big]=0,  \label{xieq}
    \end{eqnarray}
which in the UM gravity case, $w=-1$, obviously reduces to the equation on parameters of volume preserving diffeomorphisms  $\partial_\mu\mbox{\boldmath$\xi$}^\mu=0$ \cite{unimodular}.

With the decomposition of $\xi^i$ into the longitudinal and transverse parts\footnote{Since general diffeomorphism invariance is broken, the transformation properties of $\varphi$ and $\xi^i_\perp$ are no longer of a scalar and vector type, and the $\sqrt\gamma$-factor is added merely for reasons of convenience. },
    \begin{eqnarray}
    &&\mbox{\boldmath$\xi$}^i=\sqrt\gamma\,\big(\gamma^{ij}\partial_j\varphi
    +\xi^i_\perp\big),\quad
    \partial_i(\sqrt\gamma\,\xi^i_\perp)=0,
    \end{eqnarray}
the equation (\ref{xieq}) can be solved with respect to $\varphi$ in terms of the spatially nonlocal Green's function of the Laplacian operator $\varDelta$ weighted by the function $w$,
    \begin{eqnarray}
    &&\varphi=\frac1{w\varDelta}D_t\mbox{\boldmath$\xi$}^0,\quad
    \varDelta=\partial_i\gamma^{ij}\sqrt\gamma\,\partial_j,\quad
    D_t=\partial_t-(1+w)\,N^i\partial_i.
    \end{eqnarray}
The gauge parameter $\mbox{\boldmath$\xi$}^\alpha$ can be represented in terms of a projector $\mbox{\boldmath$\varPi$}^\alpha_\beta$ acting on a generic diffeomorphism parameter $\xi^\beta$
    \begin{eqnarray}
    &&\mbox{\boldmath$\xi$}^\alpha=
    \mbox{\boldmath$\varPi$}^\alpha_\beta\,\xi^\beta,\quad
    \mbox{\boldmath$\varPi$}^\alpha_\beta=
    \left[\begin{array}{cc}
    1 & 0\\
    \sqrt\gamma\, \partial^i\frac1{w\varDelta}D_t\,\, &\,\,
    \sqrt\gamma\left(\delta^i_j-\partial^i\frac1\varDelta\partial_j\right)
    \end{array}\right],
    \end{eqnarray}
so that the generators
    \begin{eqnarray}
    \mbox{\boldmath$R$}^{\mu\nu}_{\;\;\beta}=
    -2\nabla^{(\mu}\mbox{\boldmath$\varPi$}^{\nu)}_\beta  \label{generators}
    \end{eqnarray}
of the gauge invariance transformations of the action (\ref{action}) are not linearly independent. They are annihilated by the zero vector $Z^\beta_0$ of the projector $\mbox{\boldmath$\varPi$}^\alpha_\beta$,
    \begin{eqnarray}
    \mbox{\boldmath$R$}^{\mu\nu}_{\;\;\beta}Z^\beta_0=0,\quad
    Z^\beta_0=\left[\begin{array}{c}
    0\,\\
    \sqrt\gamma\,\partial^i
    \end{array}\right]     \label{reducible}
    \end{eqnarray}
Thus, this is the gauge theory with reducible generators, which should be subject to the BV technique of \cite{BV}. It is important that the generators (\ref{generators}) are nonlocal, and this would present certain difficulties in the framework of the {\em Lagrangian} quantization which is strongly based on the locality of gauge generators and structure constants. However, this nonlocality is in space rather than in time, so that time locality of the formalism is preserved and, therefore, guarantees applicability of the {\em canonical} quantization to be implemented in the future \cite{work_in_progress}.\footnote{Of course, transition from the canonical path integral to the Lagrangian one will again raise the issue of locality accompanied by the associated issues of renormalizability, etc., but this problem definitely goes beyond the present discussion of the quantization of the model.}

\section{Dynamics of dark fluid in the comoving frame}
The dynamics of the Lagrange multiplier $\lambda$ and the corresponding density and pressure is determined from the conservation law for the stress tensor (\ref{stresstensor})
    \begin{eqnarray}
    \nabla^\mu T_{\mu\nu}=
    \nabla^\mu[\,(\varepsilon+p)u_\mu u_\nu\,]+\nabla_\nu p=0, \label{cons}
    \end{eqnarray}
where in the definition of covariant derivatives we interprete $\varepsilon$ and $p$ as scalars, that is $\nabla_\nu p=\partial_\nu p$ and $\nabla_\nu \varepsilon=\partial_\nu\varepsilon$, to match with the definition of covariant derivatives acting on Einstein tensor in the l.h.s. of Einstein equation. Since the theory is not invariant with respect to general coordinate transformations the density and pressure are not scalars, and their properties are frame dependent. Three independent diffeomorphisms preserving the condition (\ref{class}) derived above are sufficient to make a transform to the distinguished comoving frame of the dark fluid. In this frame $u^i\sim g^{0i}=0$, and the temporal component of (\ref{cons}),
    $\nabla^\mu T_{\mu\nu}u^\nu\equiv-u^\mu\nabla_\mu\varepsilon-(\varepsilon+p)\nabla_\mu u^\mu=0$,
gives
    \begin{eqnarray}
    0=\frac{\dot\varepsilon}\varepsilon+(1+w)\frac{\dot\gamma}{2\gamma}=
    \partial_t\left(\ln\varepsilon+\frac12\ln\gamma+\ln N\right),
    \end{eqnarray}
where we took into account that $\nabla_\mu u^\mu=\dot\gamma/2N\gamma$ and $w\dot\gamma/2\gamma=\partial_t\ln N$. Therefore
    \begin{eqnarray}
    \varepsilon\,N\sqrt\gamma=S({\bf x}),  \label{A}
    \end{eqnarray}
where $S({\bf x})$ is a time integration constant -- some function of spatial coordinates.

Space components of the conservation law (\ref{cons}) give in the same gauge
    \begin{eqnarray}
    0=\nabla^\mu T_{\mu i}=\partial_i(w\varepsilon)
    +\frac{\partial_i N}N\,(1+w)\varepsilon,      \label{i-component}
    \end{eqnarray}
where we took into account that $u^\mu\nabla_\mu u_i=\partial_i N/N$. For the case of $w\neq 0$, dividing this equation by $w\varepsilon$  we immediately have $\partial_i(\ln w+\ln\varepsilon+\ln\gamma/2+\ln N)=0$ in virtue of the relation $\partial_i\ln N/w=\partial_i\ln\gamma/2$, so that
    \begin{eqnarray}
    w\,\varepsilon\,N\sqrt\gamma=T(t).  \label{B}
    \end{eqnarray}

Combining (\ref{A}) and (\ref{B}) together, one finds
    \begin{eqnarray}
    w=\frac{T(t)}{S({\bf x})},           \label{w1}
    \end{eqnarray}
which means that for a class of models with a constant nonvanishing $w$ both functions also degenerate to constants in space and time,
    \begin{eqnarray}
    \quad N={\rm const}\,\gamma^{w/2},\quad \varepsilon=\frac{\rm const}{\gamma^{(w+1)/2}},\quad
    w={\rm const}\neq 0.              \label{DE}
    \end{eqnarray}

For the case of the dust with zero $w$ and a constant lapse (originally considered in \cite{Burlan}) only the first term of Eq.(\ref{i-component}) remains, so that one nontrivial function of spatial coordinates $S({\bf x})$ still survives
    \begin{eqnarray}
    N={\rm const},\quad
    \varepsilon=\frac{S({\bf x})}{N\sqrt\gamma}\equiv
    \frac{\tilde S({\bf x})}{\sqrt\gamma},\quad w=0.     \label{dust}
    \end{eqnarray}

In fact, these two cases of {\em dark energy} with a constant $w$ close to $-1$ and {\em dark dust} seem to saturate physically reasonable cosmological setups in the generalized UM theory. This follows from a simple observation that a nontrivial function $S({\bf x})$ is obviously a part of initial conditions, but the parameter $w$ is determined by a {\em kinematical} restriction (\ref{class}) of the configuration space of the theory and should not depend on its particular initial conditions like (\ref{w1}) unless it is some universal constant.\footnote{Boundary conditions can in principle be incorporated into the Lagrangian as local total derivative terms forming boundary integrals in the action, which is however not the case of (\ref{class}).}

This can easily be illustrated by a simple example which shows that the attempt to model a fairly generic equation of state $p=p(\varepsilon)$
by an appropriate choice of function $N(\gamma)$ in (\ref{class}) actually fails. Consider a popular Chaplygin gas model with $p=-A/\varepsilon$. Independently of the unimodular setup, the conservation of its stress tensor, $\dot\varepsilon+(p+\varepsilon)\dot\gamma/2\gamma=0$, gives a well known relation between the energy density and $\gamma$ \cite{Chap,Chap1}, $\varepsilon=\sqrt{A+B({\bf x})/\gamma}$, where $B({\bf x})$ is a time integration constant -- some function of spatial coordinates. Together with the equations (\ref{A}) and (\ref{B}) this relation yields the expression for $N$ in terms of $\gamma$, $N=\sqrt{-S({\bf x})\,T(t)}/\sqrt\gamma$. According to the assumptions of our generalized unimodular gravity both $N$ and $w$ are the functions of one variable $\gamma$, which means that both the ratio (\ref{w1}) and the product of $S({\bf x})$ and $T(t)$ should be the functions of $\gamma$.  This is possible only when $S({\bf x})$ and $B({\bf x})$ are constant and $\gamma$ is a function of time, which means that this case, in contrast to the $w = const$ case above, is valid only for a spatially homogeneous model. Similar situation holds for other equations of state with $w \neq const$.

\section{Conclusions}

Thus, we see that there exists a class of models with a broken Lorentz invariance generalizing unimodular gravity theory, which generate dark fluid with a barotropic equation of state with a constant $w$. Similarly to UMG the gravitational dynamics of this fluid is characterized by an independent of space and time constant which is fixed by initial conditions. The spacetime rigidity of this constant implies that this fluid does not carry local degrees of freedom, but rather describes a global variable incapable of clustering. Therefore it can play the role of dark energy, especially in view of the fact that the parameter $w$ can occupy a continuous range of values near $w=-1$. For a special case of $w=0$ the rigidity condition relaxes to one constant in time function of space coordinates $S({\bf x})$, so that this {\em dark dust} can be interpreted, similarly to mimetic gravity \cite{mimetic}, as a candidate for dark matter.

 Our work, in fact, suggests a new concept in cosmology and gravity theory which can be called ``darkness" designating the general mechanism based not on local degrees of freedom, but rather on global, topological ones, that could underlie the whole bunch of phenomena and their models, including dark energy, Horava gravity theory \cite{HGrav}, quantum initial value problem \cite{SLIH,Everything}, cosmological constant sequestering formalism \cite{sequester,sequester1}, etc.

Breakdown of Lorentz invariance is perhaps too high a price for the generation of darkness phenomena in cosmology. However, Lorentz symmetry violation has become very popular in recent years due to the fact that the extension of Lifshitz anisotropic scaling invariance to gravity -- Horava gravity models -- is a way to recover unitarity in renormalizable higher derivative quantum gravity \cite{Lifshitz,HGrav}. Moreover, breakdown of Lorentz invariance can be an inalienable feature of cosmological initial conditions. The suggestion of the initial quantum state of the Universe in the form of the microcanonical density matrix \cite{SLIH} implies existence of the distinguished foliation of spacetime by spatial hypersurfaces. This foliation underlies the construction of this initial state density matrix and persists in the further cosmological evolution. Therefore, there is no reason to reject violation of Lorentz symmetry at a deeper kinematical level, like in the condition (\ref{class}).

The density matrix state \cite{SLIH} is conceptually very attractive because of the minimum set of assumptions underlying it \cite{Everything} and, moreover, because of a mechanism restricting the cosmological ensemble to subplanckian domain in UV limit and avoiding the IR catastrophe, characteristic of the no-boundary wavefunction. However, it applies only to a closed Universe with a negative contribution $\varOmega_K=-K/H^2a^2$ of the positive spatial curvature, $K=+1$, in the full set of cosmological density parameters, $\varOmega_K+\varOmega_\varLambda+\varOmega_m=1$, where $a$ is a scale factor of the FRW metric, $H=\dot a/Na$ is its Hubble factor and $K=\pm1,0$ is the sign of the 3-metric curvature scalar respectively for closed, open or spatially flat FRW cosmology. Therefore, even though the density matrix prescription generates good hill-top initial conditions for inflation (at the {\em maximum} of the inflaton potential) \cite{hill-top}, it does not include the case of a spatially flat FRW model, $K=0$, most natural from the viewpoint of the observational status of inflationary scenario ($\varOmega_K=0.000\pm 0.005$ according to combined Planck, lensing and BAO data \cite{Planck}).

Remarkably, the generalized UM model with $w=-1/3$ can imitate the effect of positive/negative spatial curvature in the Friedmann equation with $K=\pm1$, $\gamma\sim a^3$, provided the integration constant in the expression (\ref{DE}) for a dark fluid density $\varepsilon$ is negative/positive. Under a proper normalization of the flat space scale factor $a$ the dark fluid density becomes $\varepsilon_K=-3M_P^2K/a^2$ and fully imitates the spatial curvature contribution $\varOmega_K=\varepsilon_K/3M_P^2H^2$ to the {\em flat space} Friedmann equation
    \begin{eqnarray}
    H^2=\frac{\varepsilon_m+\varepsilon_\varLambda+\varepsilon_K}{3M_P^2}.
    \end{eqnarray}
This would allow one to extend the conclusions of \cite{SLIH} to FRW models in the flat and even hyperbolic space foliations, and this is one of the motivations for our generalized UM gravity model.

What underlies this phenomenon, which as we see can effectively change even the space topology \cite{Kam-Khal}, is a global degree of freedom encoded at the level of the Lagrangian formalism in the integration constant.  Like in a conventional unimodular gravity the mechanism of this is based on a subtle interplay of physical and gauge degrees of freedom -- in the generalized version it is technically more involved, but conceptually similar to the original unimodular case. A similar mechanism due to the interplay of conformal invariance and field reparametrization can be observed in the mimetic gravity theory \cite{mimetic}, though the latter incorporates a new local (dust matter) degree of freedom \cite{mimetic1}, whereas in our case this is the global topological variable parameterizing the dark fluid.

It should be emphasized that our generalized model is not a gauge fixed version of general relativity. In UM gravity the cosmological constant $\varLambda$ is incorporated as an integration constant of equations of motion and this makes a great conceptual difference from GR with a given $\varLambda$. A similar situation happens here, but the integration "constant" is much richer -- this is the perfect fluid stress tensor without local degrees of freedom.

 Here we analyzed the generalized UM gravity at the classical level.  At the quantum level its global mode should either be disentangled explicitly or treated within the quantization method for constrained systems. In either case rigorous quantization requires the construction of the canonical formalism.  As is known, UM gravity in this formalism \cite{UMcan} has instead of the GR Hamiltonian constraint the vanishing of the spatial gradient of this constraint, which eventually results in a freely chosen value of $\varLambda$ as an integration constant. As will be shown in a forthcoming paper \cite{work_in_progress}, a similar but more involved constraint appears here. At the Lagrangian level this is a conservation of perfect fluid stress tensor leading to the rigidity of its energy density and pressure, which can be interpreted as the absence of clustering of dark energy (or, in a particular case of a zero pressure, as dark matter).

 At the quantum level, especially in the transition from the canonical to the Lagrangian quantization, the situation becomes nontrivial because linear dependence of the gauge invariance generators (\ref{reducible}) implies reducibility of the first class constraints of the canonical formalism, which is subject to BV formalism for systems with linearly dependent generators \cite{BV}. Additional difficulty is that this reducibility is of a spatially nonlocal nature because of nonlocal generators (\ref{generators}).

Treatment of this problem was endeavored in \cite{UMcan,seqcan} and has lead to a special procedure of averaging over 3-dimensional space -- the counterpart to the analogous {\em spacetime} averaging in the vacuum energy sequestering mechanism of \cite{sequester,sequester1}.\footnote{It should be emphasized that this mechanism, which is an interesting part of solution of hierarchy and cosmological constant problems, can also be generalized in a Lorentz non-invariant way, what can be done by a covariantization analogous to the covariant formulation of UM gravity \cite{unimodularHT,UMcan} -- parametrization of the distinguished spacetime foliation by an auxiliary antisymmetric tensor or vector density. This, however, will have to be achieved by parameterizing all 4-dimensional coordinates in terms of four embedding functions \cite{work_in_progress}.} Weak point in this averaging procedure is an ad hoc choice of the integration measure. In particular, it fails to be well defined in noncompact asymptotically flat spacetimes. Moreover, physical predictions of \cite{sequester,sequester1,UMcan,seqcan} depend on this measure, whereas the freedom in its choice should be physically irrelevant because it reflects invariance of the BV quantization scheme under the change of the basis of gauge generators (\ref{generators}) or canonical constraints. Careful analysis of this problem will be a subject of our future work \cite{work_in_progress}. This analysis should, perhaps, resolve the conundrum of nonlocality and acausality in sequestering mechanism of \cite{sequester}, change the conclusions on spacetime compactness in the epoch of transient cosmological expansion \cite{sequester1} and, thus, extend cosmological applications to spatially open models.

\section*{Acknowledgements}
This work was supported by the RFBR grant No.17-02-00651. The work of A.O.B. was also supported by the Tomsk State University Competitiveness Improvement Program.


\begin{thebibliography}{99}
\bibitem{f(R)}
A. A. Starobinsky,  Phys. Lett. B  {\bf 91} (1980) 99;  JETP Lett. 86  (2007) 157.

\bibitem{Deser-Woodard}
S. Deser and R. P. Woodard, Phys. Rev. Lett. 99 (2007) 111301.

\bibitem{unimodular}
J. J. van der Bij, H. van Dam and Y. J. Ng,  Physica, 116A (1982) 307.

\bibitem{unimodularHT}
M. Henneaux and C. Teitelboim,  Phys. Lett. B 222 (1989) 195.

\bibitem{unimodular1}
W.~G.~Unruh,
Phys.\ Rev.\ D  40  (1989) 1048.
K. V. Kuchar, Phys. Rev. D 43 (1991) 3332;
G. F. R. Ellis, H. van Elst, J. Murugan and J. P. Uzan, Class. Quant. Grav. 28 (2011) 225007.

\bibitem{sequester}
N. Kaloper and A. Padilla, Phys. Rev. Lett. 112 (2014) 091304;  Phys. Rev. Lett. 118 (2017) 061303;
N. Kaloper, A. Padilla, D. Stefanyszyn, G. Zahariade,
Phys. Rev. Lett. 116 (2016) 051302.

\bibitem{sequester1}
N. Kaloper and A. Padilla,  Phys. Rev. D 90 (2014) 084023.

\bibitem{QCDholonomy}
A. R. Zhitnitsky,  Phys. Rev. D 89 (2014) no.6, 063529;  Phys. Rev. D 90 (2014) no.4, 043504;  Phys. Rev. D 92 (2015) no.4, 043512.

\bibitem{UMcan}
R. Bufalo, M. Oksanen, A. Tureanu,  Eur. Phys. J. C 75 (2015) 477.

\bibitem{Lifshitz}
E. M. Lifshitz,  Zh. Eksp. Teor. Fiz. 11 (1941) 255 \& 269;
E. H. Fradkin,  Front. Phys. 82 (2013) 1.

\bibitem{HGrav}
P. Horava, Phys.Rev. D 79 (2009) 084008;
A. O. Barvinsky, D. Blas, M. Herrero-Valea, S. M. Sibiryakov, and C. F. Steinwachs, Phys. Rev. D93 (2016) no. 6, 064022.

\bibitem{Venturi}
R.~Righi and G.~Venturi,
  Nuovo Cim.\ A  43  (1978) 145;
  Nuovo Cim.\ A  52  (1979) 166;
Nuovo Cim.\ A  52  (1979) 511;
Lett.\ Nuovo Cim.\   31 (1981)  487.

\bibitem{Akhmeteli}
A.~M.~Akhmeteli,
  Int.\ J.\ Quant.\ Inf.\  9  (2011) S17;
  J.\ Math.\ Phys.\   52  (2011) 082303;
 Eur.\ Phys.\ J.\ C  73 (2013)  2371.

\bibitem{Adler}
S.~L.~Adler,
  Class.\ Quant.\ Grav.\   30 (2013) 195015;
   Erratum: [Class.\ Quant.\ Grav.\   30 (2013) 239501;

  Int.\ J.\ Mod.\ Phys.\ D  25 (2016)   1643001;
Implications of a frame dependent gravitational effective action for perturbations on the Roberston-Walker Metric, arXiv:1704.00388 [gr-qc].

\bibitem{mimetic}
A.H. Chamseddine and V. Mukhanov,  JHEP 11 (2013) 135.

\bibitem{mimetic1}
 A.~Barvinsky, JCAP 1401 (2014) 014.

\bibitem{BV}
  I.~A.~Batalin and G.~A.~Vilkovisky,
  Phys.\ Lett.\  102 B (1981) 27;
  Phys.\ Lett.\   120B (1983) 166;
  Phys.\ Rev.\ D 28 (1983) 2567
  Erratum: [Phys.\ Rev.\ D 30 (1984) 508];
I.A. Batalin, E.S. Fradkin,  Lett. Nuovo Cim. 38 (1983) 393.

\bibitem{SLIH}
A. O. Barvinsky and A. Yu. Kamenshchik,  JCAP  09 (2006) 014.

\bibitem{work_in_progress}
A.Barvinsky and A.Yu.Kamenshchik, work in progress.

\bibitem{Burlan}
D.E. Burlankov, Physics of Space and Time (in Russian), Nizhny Novgorod, 2014.

\bibitem{Chap}
A.~Y.~Kamenshchik, U.~Moschella and V.~Pasquier,
  Phys.\ Lett.\ B 511 (2001) 265.

 \bibitem{Chap1}
N.~Bilic, G.~B.~Tupper and R.~D.~Viollier,
  Phys.\ Lett.\ B  535  (2002) 17.

\bibitem{Everything}
A. O. Barvinsky, Phys. Rev. Lett. 99 (2007) 071301.

\bibitem{hill-top}
A. O. Barvinsky, A. Yu. Kamenshchik and D. V. Nesterov,  JCAP 01 (2016) 036;
Eur. Phys. J. C 75 (2015) 12, 584.

\bibitem{Planck}
P. A. R. Ade et al, 	Astronomy \& Astrophysics 594 (2016)  A13 (2016).

\bibitem{Kam-Khal}
A.~Y.~Kamenshchik and I.~M.~Khalatnikov,
  Int.\ J.\ Mod.\ Phys.\ D  21 (2012) 1250004.

\bibitem{seqcan}
R. Bufalo, J. Klusoň, M. Oksanen,  Phys. Rev. D 94 (2016) 044005.

\end{thebibliography}
\end{document}